\documentclass[a4paper]{jpconf}
\usepackage{graphicx}

\def\th{\theta}
\def\ph{\phi}

\def\no{\nonumber}
\def\f{\frac}
\def\S{\Sigma}
\def\D{\Delta}
\def\Om{\Omega}
\def\c{\cite}
\begin{document}
\title{Generalized Smarr formula as a local identity for arbitrary dimensional black holes}

\author{Sujoy Kumar Modak}

\address{S.N. Bose National Centre for Basic Sciences, Block-JD, Sector-III, Salt Lake City, Kolkata-700098, India}

\ead{sujoy@bose.res.in}

\begin{abstract}
We discuss a method based on Killing symmetries and Komar conserved charges to generalize Smarr mass formula for arbitrary dimensional charged, rotating spacetime. We derive a local identity defined at the event horizon of the rotating black hole in Einstein-Maxwell gravity which reproduces the generalized Smarr formula as a by-product. The advantages of this new identity are the following: (i) unlike Smarr formula, which is non-local, this identity is purely local and hence a switchover between horizon and infinity is unnecessary and (ii) the new identity could be mapped with the recent investigations on emergent gravity. 
\end{abstract}

\section{Introduction}
Smarr mass formula \c{smarr} ensures that the energy of a (3+1) dimensional charged rotating black hole is distributed as its surface energy, electrostatic energy and rotational energy: thus supports black hole {\it no-hair} theorem. The differential form of this formula gives the first law of black hole mechanics which is analogous to first law of thermodynamics. While this mass formula is known for few decades its extension to higher dimensions considering charged-rotating spacetimes is not known. Here we first find out to what extent the existing method \c{smarr} could be useful  in order to fulfil this gap and then provide a way based on Killing symmetries and Komar conserved charges to generalize this formula to arbitrary dimensions. 

This generalized formula is shown to be a by-product of a horizon identity which relates the Komar conserved charge corresponding to a null Killing vector at horizon with its thermodynamic variables. Interestingly this identity could be regarded as the analogue of another identity associated with the timelike Killing horizon of  a spacetime which may not be a black hole.

\section{Higher dimensional spherically symmetric black hole}
In this section we use the original method of Smarr \cite{smarr} to derive mass formula for $N+1$ dimensional charged black hole represented by the Reissner-Nordstrom space-time,  
\begin{eqnarray}
ds^2=-(1-\frac{m}{r^{N-2}}+\frac{q^2}{r^{2(N-2)}})+r^2(d\theta^2 +\sin^2{\theta}d\phi^2)\nonumber\\
+r^2\cos^2\theta(d\chi_{1}^2+\sin^2\chi_1[d\chi_2^2+\sin^2\chi_2(...d\chi_{N-3}^2)] ).
\label{rn}
\end{eqnarray}
We begin by finding the position of the event horizon by setting $g^{rr}=0$, which is found to be $r_+=\left(\frac{m}{2}+\sqrt{{\frac{m^2}{4}}-q^2}\right)^{\frac{1}{N-1}}$. Using this the area of event horizon takes the form
\begin{eqnarray}
A_{H}=\frac{4\pi \bar{A}_{N-3}}{N-2}\left(\frac{m}{2}+\sqrt{\frac{m^2}{4}-q^2}\right)^{\f{N-1}{N-2}}
\end{eqnarray}
where $\bar{A}_{N-3}$ is the area of the unit sphere in $N-3$ dimensions. By inverting this relation it is possible to express mass $m$ as
\begin{eqnarray}
m=\left(\frac{(N-2)A_{H}}{4\pi \bar{A}_{N-3}}\right)^{\frac{N-2}{N-1}}+q^2\left(\frac{4\pi \bar{A}_{N-3}}{(N-2)A_{H}}\right)^{\frac{N-2}{N-1}}.\label{mrn}
\end{eqnarray}
Writing in the differential form (\ref{mrn}) takes the form $dm=\frac{(N-2)}{(N-1)\bar{A}_{N-3}}T dA_{H}+\frac{2q}{(N-2)r_+^{N-2}}dq$, where, $T=\frac{\kappa}{2\pi}$ is Hawking temperature of R-N black hole. Note that in (\ref{mrn}) we have essentially written mass in the form of a homogeneous function of degree $\frac{N-2}{N-1}$ in $A_H$ and  $q^\frac{N-2}{N-1}$ respectively. By virtue of this in build algebraic relationships among these parameters Euler's theorem on homogeneous function is readily applicable which finally extracts the mass ($m$) in terms of the variables $A_H$ and $q$ and the prefactors appearing in above differential form. Finally substituting $m$ and $q$ in terms of the physical mass ($M$) and charge ($Q$) given by $M =\frac{\bar{A}_{N-1}(N-1)}{16\pi G}m$ and $Q=\sqrt{\f{(N-2)(N-1)\bar{A}_{N-1}}{8\pi G}}q$, we find  \cite{modak2}
\begin{eqnarray}
M=Q\Phi +\f{N-1}{N-2}\f{\kappa A_{H}}{8\pi},
\label{rnsmarr}
\end{eqnarray}
where we have used the expression for electric potential $\Phi=\f{Q}{(N-2)r_+^{N-2}}$. This is the generalized Smarr formula for the metric (\ref{rn}) and we see mass of the R-N black hole could be expressed in terms of electric energy and surface energy at arbitrary dimensions. For $N=3$ it reproduces the known result. 

We must also notice that the algebraic relationship among various black hole parameters plays the key role in deriving the result (\ref{rnsmarr}). But this is not always the case. The structure of homogeneity breaks down for rotating black holes with $N>3$. Thus the above method does not go through and we need to find alternative way for deriving the mass formula for these cases. In the next section we highlight one such method (based on Killing symmetries and conserved charges) for $N+1$ dimensional charged, rotating black holes. 

\section{Higher dimensional charged, Myers-Perry black hole} 
The spacetime metric for the $N+1$ dimensional charged Myers-Perry black hole in Boyer-Lindquist type coordinates, with one spin parameter ($a$), is given by \c{aliev,dianyan}
\begin{eqnarray}
ds^2&=&-\left(1-\f{m}{r^{N-4}\Sigma}+\frac{q^2}{r^{2(N-3)}\Sigma}\right)dt^2+\f{r^{N-2}\Sigma}{\D}dr^2+\S d\th^2-\f{2a(mr^{N-2}-q^2)\sin^2\th}{r^{2(N-3)}\S}dtd\ph\no\\
&&+\left(r^2+a^2+\f{a^2(mr^{N-2}-q^2)\sin^2\th}{r^{2(N-3)}\S}\right)\sin^2\th d\ph^2+r^2\cos^2\th d\Om^2_{N-3},
\label{metric}
\end{eqnarray}
with the following identifications,
\begin{eqnarray}
\D&=&r^{N-2}(r^2+a^2)-mr^2+q^2r^{4-N},
\label{delta}\\
\S&=&r^2+a^2\cos^2\th,
\label{sigma}
\end{eqnarray}
\begin{eqnarray}
d\Om^2_{N-3}=d\chi_1^2+\sin^2\chi_1[d\chi^2_2+\sin^2\chi_2(\cdot\cdot\cdot d\chi^2_{N-3})].
\label{omega}
\end{eqnarray}
The above metric is a solution of Einstein equation only in the slowly rotating limit (small $a$) and it reproduces (\ref{mrn}) when $a=0$. For $N=3$ it provides the Kerr-Newman space-time. The metric (\ref{metric}) has two Killing vectors (timelike and spacelike) given by $\xi^{\mu}_{(t)}=(1,0,0,0\cdot\cdot\cdot)$ and $\xi^{\mu}_{(\phi)}=(0,0,0,1,\cdot\cdot\cdot)$. The corresponding conserved charges is given by the Komar definition  \cite{komar,komar2}
\begin{eqnarray}
K_{\xi^{\mu}_{(t)}}=-\frac{1}{8\pi}\int_{\partial\Sigma} \xi^{\mu;\nu}_{(t)}d^{N-1}\Sigma_{\mu\nu}
\label{meff}
\end{eqnarray}
and
\begin{eqnarray}
K_{\xi^{\mu}_{(\phi)}}=-\frac{1}{8\pi}\int_{\partial\Sigma} \xi^{\mu;\nu}_{(\phi)}d^{N-1}\Sigma_{\mu\nu}.
\label{jeff}
\end{eqnarray} 
where $\Sigma$ is a $N$ dimensional spatial volume element and $\partial\Sigma$ is its boundary defined on a $N-1$ dimensional spatial hypersurface. These expressions, when evaluated at asymptotic infinity, give mass $M$ or angular momentum $J=Ma$ up to some normalisation constant. In the next subsection we shall use these expressions and derive an identity defined at the event horizon of the space-time metric (\ref{metric}).

\subsection{The identity $K_{\chi^{\mu}}=2ST$}
Black hole event horizon is a special place where the line element $ds^2$ vanishes. In addition, for stationary cases event horizon of a black hole is also a Killing horizon where the norm of a Killing vector ($\chi^{\mu}\chi_{\mu}$) is zero. For an axisymmetric space-time like (\ref{metric}) this vector turns out to be $\chi^{\mu}=(1,0,0,\Omega_H,\cdot\cdot\cdot)$ ($\Omega_H$ is the angular velocity at horizon). This can be further decomposed as $\chi^{\mu}=\xi^{\mu}_{(t)}+\Omega_H\xi^{\mu}_{(\phi)}$ which is nothing but a linear combination of two Killing vectors. Therefore $\chi^{\mu}$ also satisfies the Killing equation $\nabla_{\mu}\chi_{\nu}+\nabla_{\nu}\chi_{\mu}=0$. It could be verified that outside the horizon this vector is a non-Killing but timelike vector. It means that  $\chi^{\mu}$ is essentially a null, hypersurface orthogonal, Killing vector at the horizon. This allows us to write down a Komar conserved charge only defined at the horizon, in the following form
\begin{eqnarray}
K_{\chi^{\mu}}=K_{\xi^{\mu}_{(t)}}+\Omega_H K_{\xi^{\mu}_{(\phi)}}.
\label{kchi}
\end{eqnarray} 

It is possible to evaluate the integrals (\ref{meff}) and (\ref{jeff}) at a finite radial distance on and outside the horizon. This constant $r$ surface is chosen to be consists of only time synchronised events \cite{cohen,cohen2,modak,modak2}. By performing the integrals (\ref{meff}) and (\ref{jeff}) we find
\begin{eqnarray}
K_{\xi^{\mu}_{(t)}}=\frac{\bar{A}_{N-3}}{G}\left(\frac{m}{2}-\frac{q^2}{r^{N-2}}-\frac{q^2a^2(a^2+r^2)}{Nr^{N+2}}~2F_{1}(2,\frac{N}{2};\frac{N}{2}+1;-\frac{a^2}{r^2})\right),
\label{meff5}
\end{eqnarray}
and
\begin{eqnarray}
K_{\xi^{\mu}_{(\ph)}} &=& -\frac{a\bar{A}_{N-3}}{4Gr^{N+4}}(K_1-K_2)\label{jeff5}\\
{\textrm{where,}}\nonumber\\
K_{1} &=& \frac{4mNr^{N+4}+2q^2r^2[a^4(N-2)^2+a^2(N-4)(N-2)r^2-4(N-1)r^4]}{N(N-2)}\\
K_{2} &=& \frac{2a^2(N-2)q^2(a^2+r^2)^2~ 2F_{1}(1,\frac{N+2}{2};\frac{N+4}{2};-\frac{a^2}{r^2})}{N+2}.
\end{eqnarray}
In the above expressions $2F_1(b,c;d;-\f{a^2}{r^2})=1+\f{b.c}{1!d}(-\f{a^2}{r^2})+\f{b(b+1).c(c+1)}{2!d(d+1)}(-\f{a^2}{r^2})^2+\cdot\cdot\cdot$ is a hypergeometric function. For $N=3$ the above expressions reproduce the effective mass and angular momentum of Kerr-Newman black hole \cite{modak}. Finally by putting $r=r_+$ in above expressions and simplifying (\ref{kchi}) we obtain the identity
\begin{eqnarray}
K_{\chi^{\mu}}=2ST,
\label{2st}
\end{eqnarray}
where $S=\frac{\pi r_+^{N-3}(r_+^2+a^2)\bar{A}_{N-3}}{N-2}$ is the Bekenstein entropy and $T=\frac{(N-4)(r_+^2+a^2)+2r_+^2-(N-2)q^2r_+^{2(3-N)}}{4\pi r_+(r_+^2+a^2)}$ is the Hawking temperature.

\subsection{Generalized Smarr formula}
Interestingly the above identity could be written in a slightly different form which could be identified as Smarr formula. To get that we reinstate the black hole parameters, $M,~J,~Q,~\Phi,~\Omega_H$ in the left hand side of (\ref{2st}) whereas the right hand side is re-expressed by using $T=\frac{\kappa}{2\pi}$ and $S=\frac{A_H}{4}$. Finally for slowly rotating limit, we obtain \cite{modak2} 
\begin{eqnarray}
M=\frac{(N-1)}{(N-2)}\Omega_H J+Q\Phi +\f{(N-1)}{(N-2)}\f{\kappa A_H}{8\pi}
\label{sm2}
\end{eqnarray}
which is the generalized Smarr mass formula for the $N+1$ dimensional charged Myers-Perry black hole.

\section{Conclusion}
We discussed a method to derive the generalized Smarr formula for arbitrary dimensional rotating, charged space-time in Einstein-Maxwell gravity. It was shown that the existing method based on Smarr's work \cite{smarr} could not be extended beyond the spherically symmetric black holes for space-time dimensions greater than four. However, using the concepts of Killing vectors and conserved charges we were able to find the identity (\ref{2st}) defined at the event horizon. This identity reproduced the Smarr formula as a byproduct. By comparing the identity with the Smarr formula we note the following differences: (i) While Smarr formula (\ref{sm2}) is dimension dependent, the identity (\ref{2st}) is dimension independent, (ii) while (\ref{sm2}) is nonlocal (as some of the parameters like $M,~J$ are defined at infinity others like $\Phi,~\Omega_H$ etc. are defined at the horizon), (\ref{2st}) is completely defined at the horizon, (iii) (\ref{2st}) could be identified with another horizon identity $E=2ST$ which is only defined at the Killing horizon, as proposed in \cite{paddy1,paddy2,paddy3} from the emergent gravity point of view. In the last case case $E$ is the Noether charge corresponding to a diffeomorphic transformation of a metric, not associated with a black hole. From this one can speculate that for black holes $E$ could be identified with the Komar conserve charge $K_{\chi^{\mu}}$ as defined in (\ref{kchi}). \\

\noindent{\it{Note added: The present affiliation of the author is IUCAA, Post Bag 4, Ganeshkhind, Pune University Campus, Pune - 411 007, India, Email: sujoy@iucaa.ernet.in .}}

\end{document}